\documentclass[
twocolumn,
superscriptaddress,
groupedaddress,
showpacs,
nofootinbib,
nobibnotes,
amsmath,amssymb, 
aps,
prl,
]{revtex4}
\usepackage{graphicx}
\usepackage{graphics}
\usepackage{dcolumn}
\usepackage{bm}

\newcommand{\In}{{\rm In}}

\begin{document}

\preprint{APS/123-QED}

\title{On a chain of fragmentation equations for duplication-mutation dynamics
in DNA sequences}

\author{M.V. Koroteev}
\affiliation{School of Biochemistry and
Cell Biology, University College Cork, Ireland} 
\email{maxim.koroteev@ucc.ie}
 
\begin{abstract}
Recent studies have revealed that for the majority of species the length distributions of duplicated sequences
in natural DNA follow a power-law tail. We
study duplication-mutation models for processes in natural DNA sequences and 
the length distributions of exact matches computed from both synthetic and natural sequences.
Here we present a hierarchy of equations for various number of exact
matches for these models. The reduction of these
equations to one equation for pairs of exact repeats is found. Quantitative
correspondence of solutions of the equation to simulations is demonstrated. 
\end{abstract}

\maketitle

\section{Introduction}

In recent years a series of duplication-mutation models related to processes
occurring in natural DNA sequences has been reported \cite{koroteev_miller2011,
koroteev_arxiv, ardnt}. The motivation for introducing these models were earlier
empirical observations on {\it length distributions} \cite{footnote1} of
identical repeats in natural DNA sequences\cite{halvak, miller_report}.
In part it was observed that when computing the length distributions within single
chromosomes or whole genome sequences these distributions tended to exhibit
power-law tails with the exponent close to $-3$\cite{gao_miller}. These observations naturally
drew attention to potential mechanisms accounting for them.

The first step for explanation of these distributions was done in
\cite{koroteev_miller2011} where empirical computational models of
chromosome evolution based on a mechanism of duplications were suggested.
The duplications in these models were thought of as random events of copying and
pasting a part of the chromosome.
If we copy a part and substitute it to another place of the chromosome,
then each such event typically results in the appearance of a pair of identical
sequences which then undergo further destruction by new duplication events and
eventually disappear but as the model generated new pairs at each time unit some balance in the number of duplicates might be expected.
It was demonstrated that this evolutionary model with random duplications
generates length distributions of {\it exact matches} or
{\it maxmers}\cite{footnote2} with power-law tails; it was also demonstrated
that the slope of these tails with the exponent $-3$ can be obtained in the
model by varying a parameter responsible for the length of the sequences which
copy-pasted at each time step: this random mechanism producing new
pairs of exact matches is further referred to as {\it source of duplications};
it is characterized by several parameters, e.g., by the length of the region for copying-pasting which is chosen in accordance
with some probability distribution. Thus, this model indicated a neutral
mechanism which generated algebraic tails in the length distributions of exact
matches and provided first qualitative explanation of the corrersponding
observations in natural genomes.
 
The models less dependent of the source of duplications but incorporating
additional mechanisms for generating heavy algebraic tails in length
distributions of exact matches were represented in \cite{ardnt, koroteev_arxiv}.
Unlike \cite{koroteev_miller2011} two basic mechanisms utilized in the models,
duplication as in \cite{koroteev_miller2011} and point mutation, reflect those in natural chromosomes.
It was demonstrated that the length distributions\cite{footnote1} of repetitive
sequences simulated by the models correspond to those observed in natural chromosomes and that the form of
those distributions also was close to algebraic with exponents of
typically around $-3$. Thus the models in question were able to reproduce these
exponents and even the amplitudes of the distributions were
fitted\cite{koroteev_arxiv} but unlike \cite{koroteev_miller2011}, the structure
of the duplication source did not influence the exponent $-3$ of length
distributions in certain parameter regime.

The important feature of the models \cite{koroteev_miller2011, ardnt,
koroteev_arxiv} was the definition of pairs of exact repeats. In
\cite{koroteev_miller2011, koroteev_arxiv} the authors used {\it supermaximal
repeats} as the basic type of exact match. Supermaximal repeats are described in
\cite{taillefer2014}; they represent a subset of exact matches with additional
conditions of maximality at the ends. On the other hand, the work \cite{ardnt}
relies on the definition of exact repeats as they are computed by {\tt mummer}
but also applies additional post-processing, imitating, to our view, the
definition of supermaximal repeats \cite{ardnt}. Nevertheless, the distinctive
feature observed for the length distributions in \cite{ardnt} was the algebraic behavior of the tails for a broad range of parameters,
while \cite{koroteev_arxiv} demonstrated that when 
mutations occurred as often as duplications (simplistically speaking), the algebraic behavior
disappeared; this point is discussed in more detail in \cite{koroteev_arxiv}.
Thus, this observation indicated that the definition of exact repeats
influence the output length distributions.

Thus, the duplication-mutation model in fact is determined by two components: a)
evolutionary mechanisms applied to the synthetic chromosome, in our case,
duplications and point substitutions and b) the definition of how to compute the
length distributions, i.e., de facto, how we count exact matches. 

In
this paper we 1) rely on {\tt mummer} in our computation of the exact repeats
following \cite{ardnt} but {\it do not apply additional postprocessing} to
portrey supermaximal repeats, thus, our counting is different both from
\cite{koroteev_arxiv} and \cite{ardnt}; 2) suggest dynamic equations reproducing
both the exponent and the amplitude of the length distribution for that
counting; 3) demonstrate that the
stationary equation that we derived, reproducing the amplitude and the exponent for length distributions 
of pairs of exact repeats can be represented as a (infinite) sum or a chain of
equations for different types of exact repeats; 4) demonstrate that the equation
for supermaximal repeats from \cite{koroteev_arxiv} is incorporated in the chain of equations
we introduce for various types of exact matches.

\section{Model}

The evolutionary mechanisms used in numerical simulations of the model
correspond to \cite{koroteev_miller2011, koroteev_arxiv}: a detailed
explanation of these duplication-mutation models can
be found, e.g., in \cite{koroteev_arxiv} but we summarize them in this section.
 
\begin{figure} 
\includegraphics[width=240pt, height=160pt]{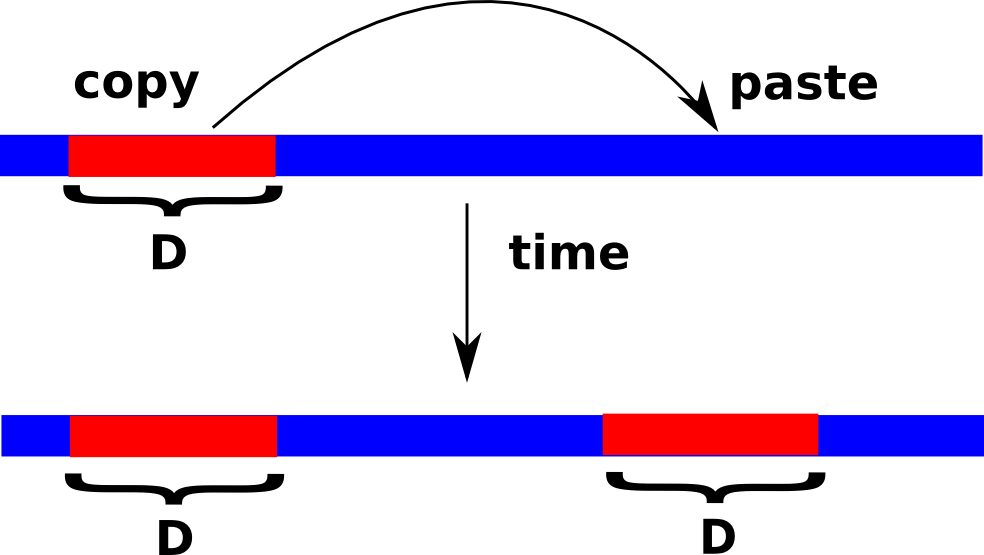}   
\caption{The figure represents random duplications as they appear in the
synthetic sequence. A random sequence
of the fixed length $D$ (red bar) is chosen from the chromosome (blue bar) and
copied into another randomly chosen place of the chromosome thus producing a pair of exact matches.
Simultaniously point substitutions are applied to the whole
chromosome with some rate. Length distributons of such pairs (with
restrictions layed by {\tt mummer}) is computed and analyzed throughout the
paper.}
\label{fig:layout}
\end{figure}

The layout of the model is shown in fig. \ref{fig:layout}.
We consider a synthetic chromosome (blue bar in fig. \ref{fig:layout})
represented as a string of $L$ bases chosen from a finite alphabet; in natural genomes the alphabet consists of 
four bases A, G, C, and T. The distance between bases is a length scale
denoted by $a$; for natural genomes it is close to $1\AA$. 

Within our models a subsequence of length $D$ (red bar in fig. \ref{fig:layout})
is chosen randomly within the chromosome and is substituted for a sequence of length $D$ at another randomly
chosen position in the chromosome (fig. \ref{fig:layout}). These duplications
are assumed to occur with the rate $\lambda$ measured per time unit, per base.
Simultaniously point substitutions are applied to the system with the rate
$\mu$ per time unit, per base.
  
The 
sequence feature that we study is the set of repeated sequences within the
chromosome. For finding all pairs of exact matches in the synthetic sequence we
apply {\tt mummer}. Mummer searches for maximal repeats or {\it maxmers}\cite{footnote2} 
which are akin to supermaxmers\cite{footnote3-1} mentioned in the previos
section and used in \cite{koroteev_arxiv} in the sense that computation of both
sets is based on some maximality condition.
However, the set of exact matches computed by {\tt mummer} is larger than the set of supermaxmers of the same length as the definition of
the latter includes additional restrictions.
Then the observations show that the output of these
computations is noticeably different if we compare the length distributions obtained in the models \cite{ardnt} and \cite{koroteev_arxiv}. 
Our aim here is the model capable to reproduce the simulated length distributions 
obtained with {\it mummer} without any additional restriction as well as an
equation for the simulated length distributions.
In the discussion below it is always implied that {\it mummer} is used with the option {\it -maxmatch} 
which according to the {\it mummer} manual produces
computations of exact matches \lq regardless of their
uniqueness\rq\cite{mummer_manual}. The Appendix section also contains more
rigorous definitions of various types of repeats. However for the purposes of
the analytic derivation suggested below it is sufficient to think that the
equations aim to reproduce the length distributions constructed for the set of
repeats obtained by {\tt mummer}, a standard tool in comparative analysis of
long DNA.

\section{Analytic treatment}

Let the number of pairs of duplicates of the length $m$ at time moment $t$ is
$g_{2}(t,m)$. We assume that new duplication events occur with the rate
$\lambda$ per base, per time unit; at the same time the chromosome undergoes
point mutation events occurring with the rate $\mu$ per base, per time
unit. We first write down the evolutionary (balance) equation for the
average number of {\it pairs of duplicates} $g_{2}$, which was derived in
\cite{koroteev_arxiv}; it has the form 
$$ 
\frac{\Delta g_{2}}{\Delta t}= - 2\left[(m+D-a)\frac{a\lambda}{D}+\mu m\right]g_{2}(t,m) + 
$$
\begin{equation}
\label{pm}
+4\left(\frac{a^{2}\lambda}{D} + a\mu\right)\sum_{k=m+1}^{D}g_{2}(t,k) +
L\frac{a\lambda}{D}\delta_{c}(D-m). 
\end{equation}
The main difference between this equation and the equation of
\cite{koroteev_arxiv} is notation (we use $g_{2}$ here instead of $f$). In
addition, there is no prefactor $2$ in the last term of the equation because in \cite{koroteev_arxiv} we studied the number of duplicated {\it sequences} while
here we look at the number of {\it pairs of duplicates}; thus, the source
produces one pair of duplicates at each time step. We also
confine ourselves to the equation for the monoscale source using Kronecker
delta function $\delta_{c}(D-m)$; different source terms are also possible and
will be presented elsewhere. Thus the equation (\ref{pm}) is provided for the
reference and connection to the subsequent discussion.

We will then focus on the stationary version of the equation implying that when
$t\to\infty$ $g_{2}(t,m)\to g_{2}(m)$ (this can be demonstrated by analytic
calculation)
$$
0= -
2\left[(m+D-a)\frac{a\lambda}{D}+\mu m\right]g_{2}(m) + 
$$
\begin{equation}
\label{duplets}
+4\left(\frac{a^{2}\lambda}{D} + a\mu\right)\sum_{k=m+1}^{D}g_{2}(k) +
L\frac{a\lambda}{D}\delta_{c}(D-m). 
\end{equation}
Now in the same way as we looked at {\it pairs of identical duplicates} we can
look at triplets, quadruplets, etc.
of identical sequences and write down
the corresponding equations for them.
For $i$-plets we will have the following stationary equation
\begin{widetext}
$$
0= -
i\left[(m+D-a)\frac{a\lambda}{D}+\mu m\right]g_{i}(m) + 
2i\left(\frac{a^{2}\lambda}{D} + a\mu\right)\sum_{k=m+1}^{D}g_{i}(k) 
$$
\begin{equation}
+(i-1)\left(\frac{a\lambda}{D}(D-m+a)\right)g_{i-1}(m) +
2(i-1)\frac{a^{2}\lambda}{D}\sum_{k=m+1}^{D}g_{i-1}(k), \; i>2
\label{iplets}
\end{equation}
\end{widetext}
We see that unlike the equation for duplicates containing the source term with
the delta function in it, other equations also have sources of new $i$-plets
; these sources are
$i-1$-plets and expressed by the last two terms in (\ref{iplets}).
One produces $i$-plicates of $i-1$-plicates of the same length $m$ (the first
term in the second line of (\ref{iplets})); the other generates $i$-plicates of
longer $i-1$-plicates by copying and pasting their parts of the length $m$
(the second term in the second line of (\ref{iplets})), i.e., new duplicates,
$g_{2}(m)$ generated by the source, in turn produce triplicates
$g_{3}(\tilde{m})$, where $\tilde{m}\le m$, triplicates produce quadruplicates
$g_{4}$ etc.
The first term in the first line of (\ref{iplets}) is  responsible for the destruction of sequences by new
duplications and point mutations; coefficients represent the corresponding
rates. The second term in the first line of (\ref{iplets}) shows that longer
sequences are turned into shorter ones, again, by duplications and point
mutations. The general mechanism has much in common with models
studied in fragmentation theory\cite{ben-naim}. This similarity is also
discussed below.

Thus for each $m=1\ldots D$ we
have a set of equations for various sets of identical repeats (maxmers). 
As it was demonstrated in \cite{koroteev_arxiv} the equation for $g_{2}$ fits
well to the length distribution of supermaxmers computed for the synthetic
chromosome after applying evolutionary duplication-mutation dynamics described
above. Equations for different types of repeats, to our knowledge, were not
obtained earlier. We refer to this set of equations as {\it chain} because
as it is easily seen functions $g_{i}$ represented in the $i$-equation are
related to the ``adjacent'' functions $g_{i-1}$ and $g_{i+1}$.

Using these equations we can obtain the equation corresponding to the length
distributions of exact matches computed by {\tt mummer} as follows. We sum
up all the equations for $g_{i}$, $i=1,2,\ldots$ and find a new equation for the
function $G(m)=\sum ig_{i}(m)$; the equation has the form 
$$ -(\zeta+2)mG(m) + 2aG(m) + 2(\zeta+2)a\sum_{n>m}G(n)+ $$
\begin{equation}
+ L\delta_{c}(D-m)=0,
\label{maxmers}
\end{equation}
where $\zeta=D\mu/a\lambda$ is a dimensionless parameter.

\begin{figure} 
\includegraphics[width=240pt, height=160pt]{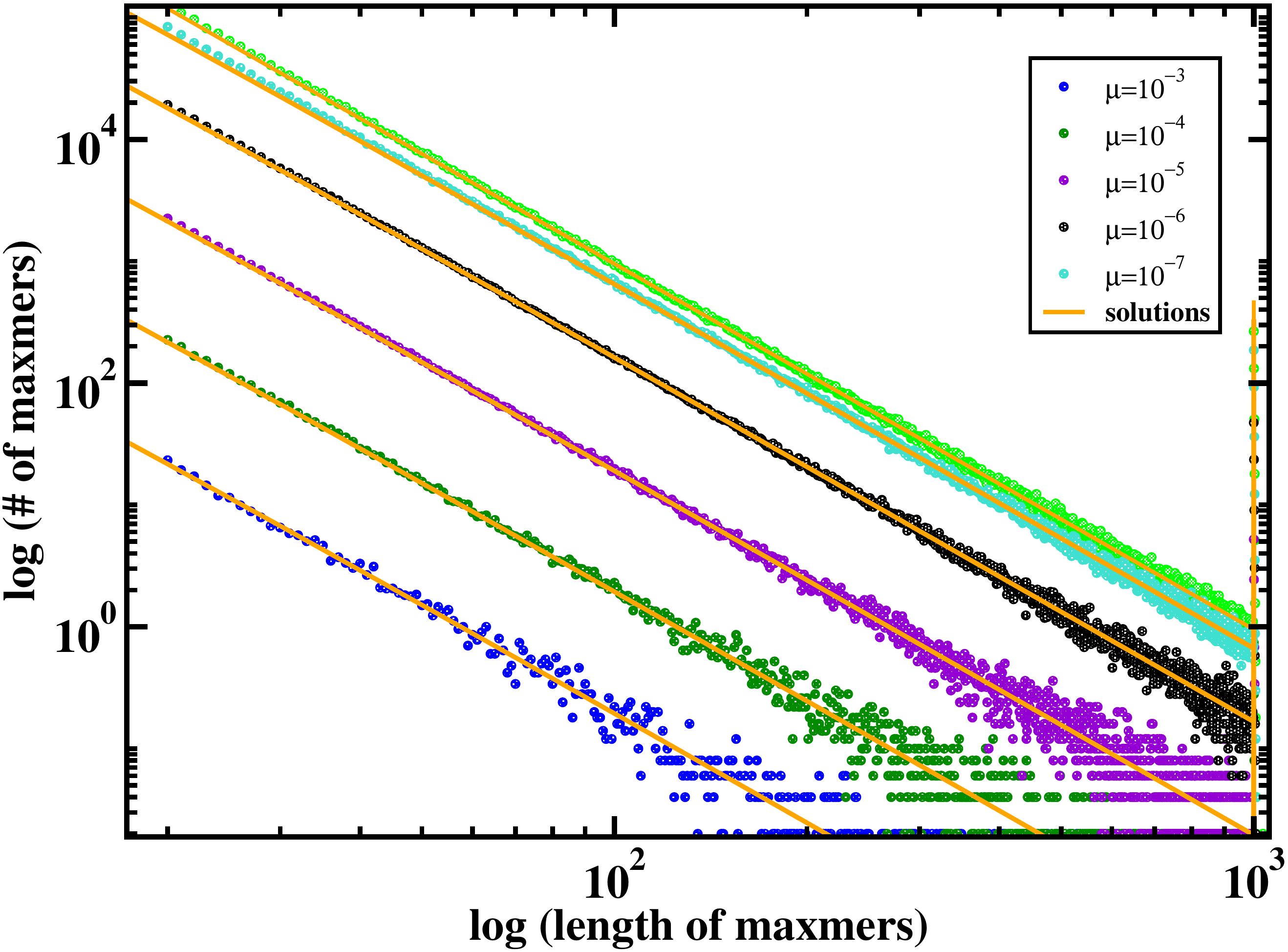}   
\caption{Curves represent stationary length distributions obtained from
simulations of duplication-mutation dynamics described
in the previous section with a monoscale source for various base substitution
rates $\mu$ and corresponding analytic solutions (orange) of (\ref{maxmers}).
The chromosome length $L=10^{6}$; source length $D=10^{3}$, duplication rate
$\lambda=10^{-4}$; for simulations we always take $a=1$.  Length distributions 
for the same dynamics computed by mummer\cite{mummer} were obtained using
the following options {\it -maxmatch -n -b -l 20}. The results were then
averaged over $10^{2}$ realizations.}
\label{fig:fig1}
\end{figure}

Now we can compare the results of the simulations with the solutions of
(\ref{maxmers}); the comparison is
represented in fig. \ref{fig:fig1}.
Additional
comparisons for different sets of parameters are given in supplemental figures
(see Supplemental materials). 
\begin{figure}
\includegraphics[width=240pt, height=160pt]{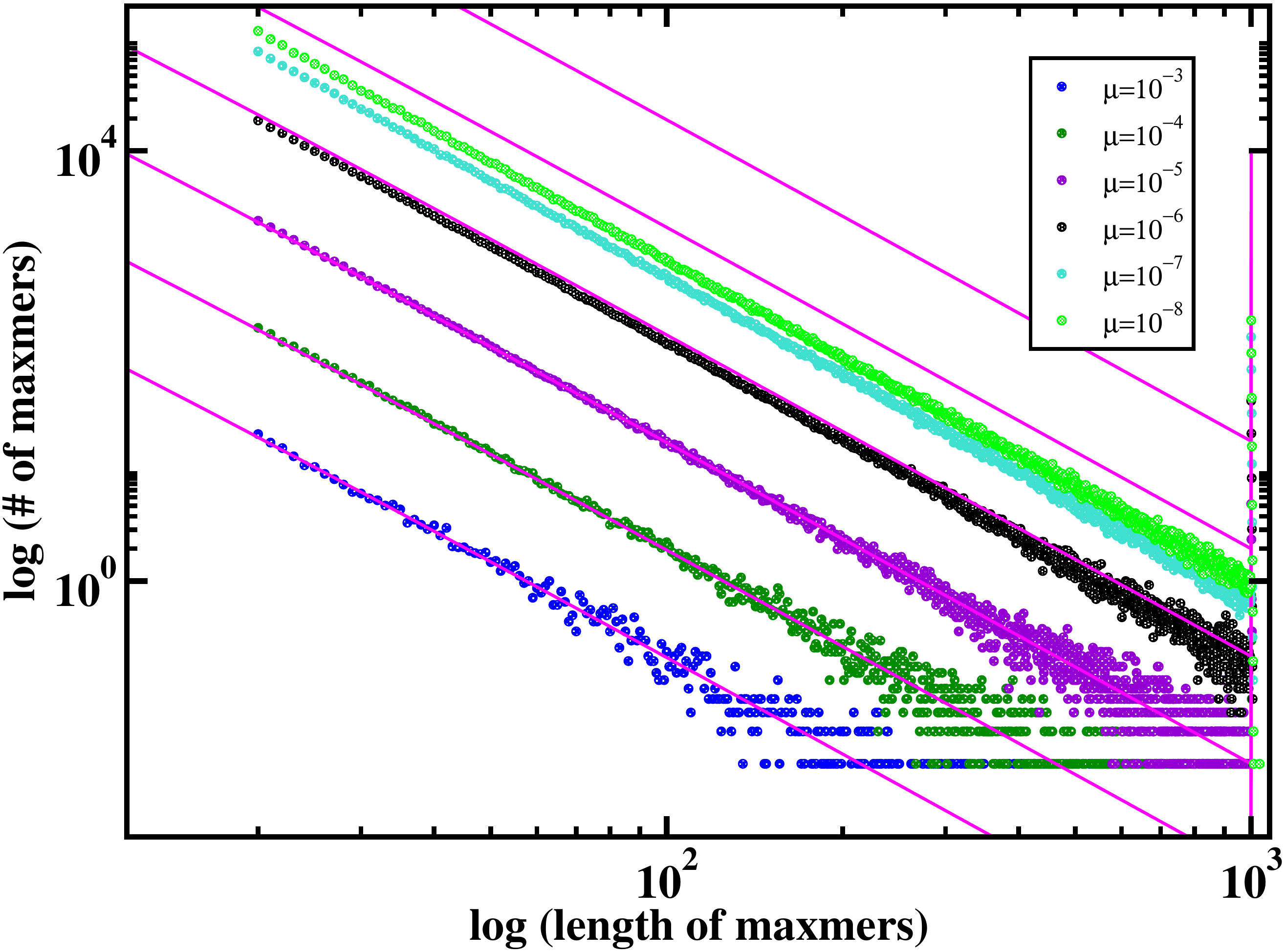}   
\caption{Curves represent stationary length distributions obtained from
simulations of duplication-mutation dynamics with a monoscale source for various base substitution
rates $\mu$ and corresponding analytic solutions (magenta curves) of eq. (5)
of \cite{ardnt}. All parameters for the simulations and the equation are the same as
for fig. \ref{fig:fig1}. The results of simulations were
averaged over $10^{2}$ realizations.}
\label{fig:fig3}
\end{figure}
Let us now compare solutions of the equation presented in \cite{ardnt}
with the simulations of the same duplication-mutation dynamics.
For that we used equation (5) of supplemental materials of \cite{ardnt}. 
Comparisons are represented in fig. \ref{fig:fig3}. 
The solutions
of \cite{ardnt} provide a good agreement for sufficiently large mutation rates
compared to the duplication rate $\lambda$ but fail to reproduce the amplitude of
the length distributions for different regimes. In this regime 
saturation is observed wrt. the amplitude of the length distributions which is reproduced by
solutions (\ref{maxmers}) as seen in fig. \ref{fig:fig1} and supplemental
figures 1 and 2\cite{footnote6}.

One then can easily understand the qualitative correspondence of length 
distributions observed in \cite{ardnt} and \cite{koroteev_arxiv} for high
mutation rates:
the growth of mutation rate $\mu$ evidently affects $g_{i}(m)$ for larger $i$ as the growth
of $i$ means {\it more sequences} in the set which are destroyed faster
affected by mutations.
Thus the main contribution to $G(m)$ for high mutation rates comes from
$g_{2}(m)$ , i.e., $G(m)\sim g_{2}(m)$ as $\zeta\to\infty$ and the dynamics is described by
(\ref{duplets}) in the main order.
Also it is instructive to note that the situation $\mu\gg\lambda$ generally implies $\zeta\gg 1$ and one can neglect in (\ref{maxmers}) all terms compared to those
containing $\zeta$ and the source term with delta function to keep the algebraic
tail, hence $L/a$ has to grow as $\sim\zeta$ to keep the same order of the
source term $\delta_{c}(D-m)$, otherwise the tail disappears as it is seen from
fig. \ref{fig:fig1} for large $\mu$: here $\zeta$ is growing but the length $L$
remains fixed.
However this is not applicable even for $\zeta\sim 1$. On the other hand, if
$\mu\ll\lambda$ then $\zeta\to 0$ and we can write down the equation corresponding to the limit of
absent mutations as $\zeta$ becomes negligible compared to $1$.
\begin{equation}
-2mG(m) + 2aG(m) + 4a\sum_{n>m}G(n)+ L\delta_{c}(D-m)=0.
\label{nomut}
\end{equation}
If $D$ is fixed as in figs. \ref{fig:fig1}, \ref{fig:fig3}, then the limit
amplitude of the algebraic tail is controlled by the only parameter $L$ and 
all distributions with decreasing $\zeta$ asymptotically have the
saturation line; this line establishes an upper boundary for fitting the model to the natural
sequence. This also can be seen from the exact solution of (\ref{nomut}) that
has the form
$$
G(m) = \left\{
\begin{aligned}
\frac{aDL}{(m-a)m(m+a)},\; & m<D \\
\frac{L}{2(D-a)},\; & m=D
\end{aligned}
\right.
$$
with obvious main order term $\sim 1/m^{3}$ as $a\ll m$. The solution is
applicable if $a\ll D\ll L$; otherwise finite size effects turn out to be
strong.

The existence of saturation also can be viewed from the continuum limit of the
dynamics under consideration. Introducing dimensionless variables 
$$
\bar{a}=\frac{a}{D}, \; \bar{m}=\frac{m}{D}, \; \bar{L}=\frac{L}{D},
$$
so that $D$ corresponds to $1$, we see that the dimensionless size of
the lattice $\bar{a}\ll 1$ and hence $\bar{a}\to 0$. We then denote $\bar{m}=x$
and taking into account that $L/D \gg 1$, we also take $\bar{L}\to\infty$;
other parameters may vary.
Then $\bar{L}\delta_{c}(1-x)$ turns into Dirac delta and the equation (\ref{maxmers})
takes the form
$$
-(\zeta+2)xG(x) + 2(\zeta+2)\int_{x}^{\infty}G(y)dy + \delta(1-x) = 0.
$$
This equation corresponds to the stationary form of eq. (1) in \cite{ben-naim}.
Its solution is
\begin{equation}
G(x) = \frac{1}{\zeta+2}\left[\frac{\delta(1-x)}{x} + \frac{2}{x^{3}}\right].
\label{cont_sol}
\end{equation}
The function has the exponent $-3$ for all $x\in (0,1)$. It is seen that the
apmplitude of the distribution $G(x)$ is controlled by the parameter
$1/(\zeta+2)$, while the slope remains the same, but in new variables $\zeta$
has the form $\mu/\lambda\bar{a}$ and as in the continuum limit $\bar{a}\to 0$
the tail $-3$ vanishes unless at least $\mu/\lambda\sim\bar{a}$. For small
$\zeta$ the dependence of the amplitude on the parameters $\mu$ and $\lambda$
disappears which corresponds to the observed saturation.

\section{Comparison to natural data}

\begin{figure}
\includegraphics[width=240pt, height=160pt]{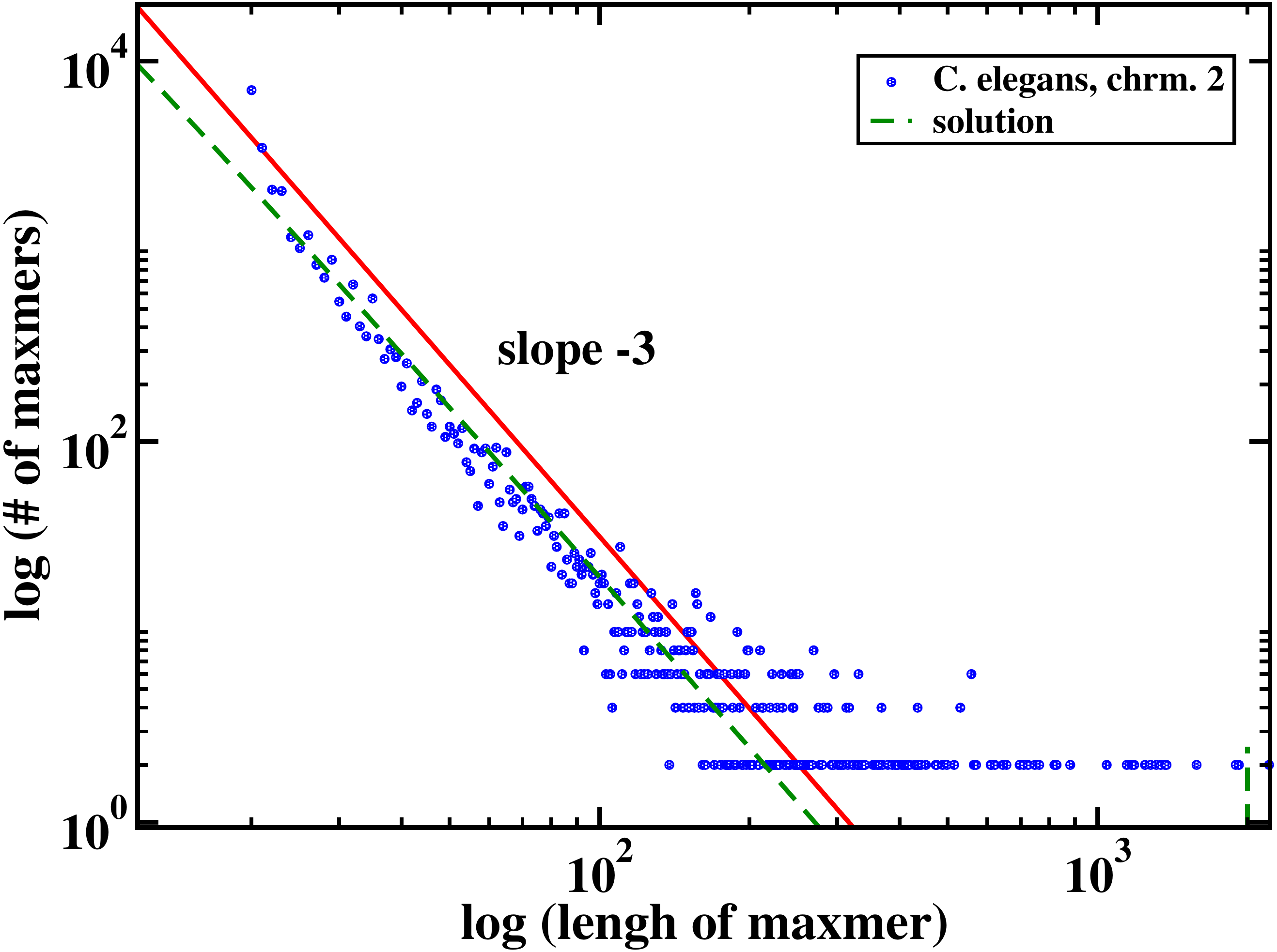}   
\caption{The length distribution for repeat-masked {\it C. elegans} chromosome $2$ was computed using {\it mummer} with the
options {\it -maxmatch -n -b -l 20};
self-hits were removed from the distribution. The length of the chromosome is $\sim 10^{7}$.
The dotted curve represents the solution of eq. (\ref{maxmers}) for the
parameters computed for the natural chromosome $D=2000$, $\mu=2\times 10^{-2}$,
and $\lambda=2\times 10^{-2}$.}
\label{fig:fig4}
\end{figure}

For the comparison of our results with natural data we take {\it C. elegans chromosome 2}, for which 
we show the length distribution of exact matches on fig. \ref{fig:fig4}.
As all synthetic sequences when processed with {\it mummer} do not contain
``self-hits'', i.e., identical sequences located exactly in the same positions
for both copies of the chromosome, the self-hits were also removed from the {\it
mummer} output for the natural sequence. To estimate the parameters of our model for this chromosome we use the estimate
for the duplication rate $0.0208$ per gene, per $1${\it my}(million
years) or $\approx 400$ duplications occur in genes per $1${\it
my}\cite{lynch}, as the number of genes in the {\it C. elegans} genome is
estimated to be around $2\times 10^{4}$\cite{choi}, or $\beta_{0}=40$ per
$1${\it my} for chromosome $2$ of length $\sim 10^{7}$ bases (as the length of
the whole genome is taken to be $\sim 10^{8}$ bases); for the rate per base
$\lambda_{0}$ we have $\beta_{0}/L_{0}$, where $L_{0}$ are bases in the {\it C. elegans} chromosome $2$ belonging to genes. 
It is known that genes cover around $50\%$ of the whole genome in C. elegans,
hence $L_{0}\approx 5\times 10^{6}$.  We assume that the duplication rate for
non-coding parts of the chromosome $\lambda=\lambda_{0}\sim 10^{-5}$ per
base, per $1${\it my}.
Then we find that $\lambda L=100$ duplications occur in coding and non-coding
parts of {\it C. elegans} chromosome 2 per $1${\it my}.

For the mutation rate in {\it C. elegans} we accept the estimate $\approx
2\times 10^{-2}$ per base, per $1${\it my}\cite{drake}; one generation $=$ four
days.
To map the parameters of the natural chromosome to the model we use the estimate for the algebraic tail of the length distribution
$2D^{2}\lambda/(\mu m^{3})$. This estimate follows from the
prefactor in (\ref{cont_sol}) if we take into account that $x\approx\bar{m}=m/D$
and $a=1$. The amplitude of the distribution for any specific $m$ is estimated
directly from the plot. In addition, it is necessary to take
into account that $\lambda=\lambda_{model}$ from  (\ref{cont_sol}) is related to
the duplication rate in the natural chromosome $\lambda_{nat}=10^{-5}$ as
$\lambda_{model}=D\lambda_{nat}/a$.
From all previous estimates we obtain $D\approx 2000$ and
$\lambda_{model}=2\times 10^{-2}$.
These estimates yield the solution of eq. (\ref{maxmers}) shown in fig.
\ref{fig:fig4}. The exact matches of the length $>200$ observed in the fig.
\ref{fig:fig4} imply that the realistic source of duplications
should have non-zero variance unlike the delta source studied here. However, as
it was shown in \cite{koroteev_miller2011}, such source does not influence the
form of the tail for length distribution.


\section{Discussion}

The solutions of the duplication-mutation dynamics presented in the paper
raise a number of questions. For the explanation of heavy algebraic tails
observed in length distributions of natural sequences we used the solutions of
the equations for $t\to\infty$. In connection with biology it should not be
unserstood as an effort to say that natural sequences are in fact in a
stationary state. First, the models studied here include only two processes
having some analogies with processes in natural DNA. Therefore it would not be
correct to interpret them as the models of how natural sequences have been varying in
their history {\it de facto}. For example, in \cite{koroteev_miller2011} we
demonstrated that long range correlations detected in natural DNA were not found in the synthetic
sequences obtained by means of these models; i.e., the length distributions
merely reflect some important evolutionary features of natural DNA neglecting
other features. Second, it is necessary to stress that basic assumptions of the
model imply uniform mutation and duplication rates both in time and in space
while in natural genomes these quantities may vary depending, e.g., on the
function of a DNA region. Nevertheless the correspondence of the solutions
to the model and natural data demonstrates that the equations detect essential
details of the data. On the other hand, it is hardly possible to indicate a
characteristic time scale for all eukariotic sequences on which significant evolutionary changes occurred to form the modern genomes. Therefore,
as the time for natural sequences is restricted by the present moment, we do not
have sufficient evidence to map this time moment to a specific time moment of
the model and the most plausible assumption is to map it to the stationary state
of the model attained for $t\to\infty$ (in the units of the model).
This assumption is confirmed by observations that stationary length
distributions of the model reproduce the length distributions of
natural sequences. However, this should be rather understood as a sojourn of a
non-stationary solution in the neighbourhood of the stationary 
one sufficiently long time compared to a characteristic time scale in
the system rather than a ``fixation" of natural genomes in stationary states and
thus the stationary system approximates well the natural DNA while the latter
still may remain non-stationary. Obviously, if a natural chromosome demonstrates
noticeable deviations from algebraic tail or other deviations from stationary solution, the assumption of non-stationarity
becomes possible and has to be studied separately. 

The equations (\ref{maxmers}) have several features deserving to be stressed.
First of all, the equations we derived for $G(m)$ allow the length distributions of exact matches computed by mummer
in a broad range of parameters to be reproduced correctly. That means, in part, that histograms computed by
counting pairs of maximal exact matches with {\tt mummer} can be understood as
$\sum ig_{i}(m)$ i.e., they represent a cummulative sum of all sequences of
duplicates, triplicates, etc.
It is worth noting that the mummer output does not compute functions $g_{i}(m)$
directly and thus the question of interpretation of $g_{i}$ in terms of
biologically meaningful sequences remains open: we observe only some
cumulative effect of distributions for $g_{i}(m)$.
On the other hand, the correspondence of functions $g_{2}(m)$ 
to the length distribution of {\it supermaxmers}
indicates a potential way to resolve this issue: if functions $g_{2}(m)$ were
interpreted as supermaxmers then the candidates for $g_{3}(m)$, $g_{4}(m)$ etc.
could be so called \lq local maxmers\rq\cite{taillefer2014,footnote5}.
At the same time the observed correspondence of {\tt mummer} output and the
function $G(m)$ suggests we have an analytic interpretation for the length
distributions computed by mummer for natural sequences: the length distributions for natural sequences
exhibiting algebraic behaviour with the exponent $-3$ can be understood in terms
of equations (\ref{iplets}) and (\ref{maxmers}) and their solutions.

The representation $G(m)=\sum_{i}ig_{i}(m)$ also indicates that the function
$G(m)$ for each $m$ can be thought of as average number of sequences $\bar{i}$
if $g_{i}(m)$ implies a non-normalized distribution function of the number
of sequences per one exact match over $i$.
The equation (\ref{maxmers}) has the form of a fragmentation equation with an input 
and thus can be construed as stationary fragmentation equation of these average
quantities $G(m)$.

We also proposed a hierarchy of equations for $g_{i}$; the first of these
equations, i.e, for $g_{2}$, was derived in \cite{koroteev_arxiv} and we see that the equations of \cite{ardnt} and \cite{koroteev_arxiv} as
well as those presented here treat different subjects focusing on various restrictions imposed on exact matches; in part, the work in
\cite{koroteev_arxiv} deals with the collection of \lq supermaxmers\rq, specific
pairs of exact repeats computed with additional conditions of maximality which are discussed
in\cite{taillefer2014}(see Appendix 1); they are important as the equations for
them not only account for the observed algebraic behaviour in length distributions of natural
DNA sequences but demonstrate, in part, non-algebraic length distributions also
observed both in simulations and natural DNA and also because their definition
provides them with a natural biological interpretation\cite{taillefer2014}.
They are accounted for by equation (\ref{duplets}) and demonstrate obvious discrepancy
from the length distribution of exact matches (suppl. fig. 3). 
Our equation (\ref{maxmers}) treats all pairs of exact matches neglecting
their uniqueness and reproduces their length distributions. Then $G(m)$ in our interpretation may be represented as a sum of \lq
supermaxmers \rq for which the biological interpretation was already discussed and
other sets of sequences obtained by natural extension of the concept of
supermamxers; in this sense, we expect that such an
interpretation of $g_{i}(m)$, $m>2$ will appear soon.

The author is acknowledged to Kun Gao for helpful discussion.

\section{Appendix 1. To the definition of excat matches}

In the appendix we provide more rigorous definitions of maximal repeats or
matches which were used in the paper but which allow to distinguish 
the results presented here from those obtained earlier. There may be several
approaches to the definition of exact matches and supermaximal repeats (cf.
\cite{taillefer2014}); our approach construes the sequence as a {\it set} and
thus all definitions are given in terms of sets and subsets.

\noindent
{\bf\S 1.} Consider a finite sequence of objects $x_{i}$, $i=1,2\ldots L$, 
$L<\infty$. For each element of the sequence there is a pair $\{i, x_{i}\}$, where
$i$ is the number of an element in the sequence\footnote{we use this redundant
notation only for clarity. It is clear that notation $\{x_{i}\}$ is enough to
denote the {\it set} of pairs, thus below $y_{k}$ may again denote the set of
$k$ pairs $\{k, y_{k}\}$}; hence, we have {\it a set } of pairs
$\{i,x_{i}\}_{i=1}^{L}$. We denote this set by $X$. By $X_{k}$ we denote a subset of $X$ consisting of $k$ pairs 
$\{i,x_{i}\}$corresponding to $k$ {\it consecutive} elements of the sequence. In
the case of DNA sequences the sequence of the length $L$ corresponds to the
whole chromosome, or whole genome or even any long DNA sequence.

\noindent
{\bf\S 2.} The configuration
space is defined by possible values of $x_{i}$. In general situation we can
assume that this space $S$ is the same for all sites of the sequence and $S=\{0,1,2,\ldots,N-1\}$. Thus, we have
$N^{L}$ possible states of the system. Consider
also the set $Y$ of all arbitrary $N$-ary sequences containing $1\le l\le
L$ elements. This is a finite set with the cardinal number $\left|Y\right| =
\sum_{k=1}^{L}N^{k}=N(N^{L}-1)/(N-1)$. Elements of this set will be
denoted by $y_{k}$ where index $k$ implies the number of elements 
the corresponding sequence. The elements of $y_{k}$ are denoted
$y_{k}=(y^{1}_{k},y^{2}_{k},\ldots,y^{k}_{k})$. For DNA sequences the
configuration space has the form $S=\{A, C, G, T\}$.

{\bf Example}. Let the configuration space be binary, i.e., $S=\{0,1\}$.
Consider the sequence $\mathfrak{X}=\{10101010\}$ for which $L=8$. The {\it set}
$X$ is represented as follows
$$
X=\{\{1,1\},\{2,0\},\{3,1\},\{4,0\},\{5,1\},\{6,0\},\{7,1\},\{8,0\}\}.
$$
For this set one of the $X_{3}$s is given by $\{\{2,0\},\{3,1\},\{4,0\}\}$. The
set $Y$ consists of all binary sequences containing  $l$ elements, $1\leq
l\leq 8$. An example of an arbitrary $y_{4}$ is furnished by an arbitrary
binary sequence of $4$ elements.

\noindent
{\bf\S 3.}
We say that the element $y_{k}\in Y$ {\it intersects} with the sequence $X$ if
$\exists a: 1\leq a \leq L-k$ such that $y^{j}_{k}=x_{a+j-1}$, $j=1,2,\ldots k$.
In our example the element $y_{4}=\{1010\}$ intersects with $X$ three times.
The subsets of $X$ corresponding to these intersections are given by
$X^{1}_{4}=\{\{1,1\},\{2,0\},\{3,1\},\{4,0\}\}$,
$X^{2}_{4}=\{\{3,1\},\{4,0\},\{5,1\},\{6,0\}\}$,
$X^{3}_{4}=\{\{5,1\},\{6,0\},\{7,1\},\{8,0\}\}$.

Let the element $y_{k}\in Y$ intersected with $X$ and the intersection is given
by the set $\{X^{1}_{k},X^{2}_{k},\ldots X^{r}_{k}\}$. We denote that by
$y_{k}=\{X^{1}_{k},X^{2}_{k}\ldots X^{r}_{k}\}$ where $X^{j}_{k}\subset X,
\forall j$. 

{\bf Definition 1}. The element $y_{k}=\{X^{1}_{k},X^{2}_{k}\ldots
X^{h}_{k}\}\in Y$ is referred to as
{\it sub-maximal k-mer} if $h>1$.

{\bf Definition 1}$^{\prime}$. Each pair of sets $(X^{i}_{k},X^{j}_{k})$, $i\ne
j$ of $y_{k}$ is referred to as {\it exact match}.

{\bf Definition 2}. Exact match $(X^{i}_{k},X^{j}_{k})$, $i\ne
j$ is referred to as {\it maximal exact match} if
at least one of $X^{i}_{k}, X^{j}_{k}\notin X^{s}_{k+p}$ $\forall p\geq 1$
and $\forall s$ such that $X^{s}_{k+p}\in
y_{k+p}=\{X^{1}_{k+p},\ldots X^{b}_{k+p}\}$ where $y_{k+p}$ is a sub-maximal
$k+p$-mer.

{\bf Example.} 

Consider the sequence
$$
TGGT\underline{GGTTA}ATTCACA\underline{GGTTA}CA\underline{GGTTA}GGG
$$
Its subsequence $GGTTA$ is a sub-maximal $5$-mer with $h=3$. Each pair of three
sequences of it forms an exact match. On the other hand, a maximal exact match
is formed by any pair except that, containing the sequences $2$ and $3$ as both these
sequences turn out to be immersed into longer sub-maximal maxmer $ACAGGTTA$.
This can be expressed in other words by saying that maximal exact matches can
not be extended even by one symbols to the left or to the right to remain in the
same time exact matches.

\noindent
{\bf \S 4.}
 For further purposes we should notice that a sub-maximal $k$-mer can be
contained into another submaximal $k+p$-mer,  $p>0$ in the sense that it may
occur that $\forall$ $X^{i}_{k}$ there exists $X^{j}_{k+p}$: $X^{i}_{k}\subset
X^{j}_{k+p}$. This observation motivates the following definition.

{\bf Definition 3}. The sub-maximal $k$-mer $y_{k}=\{X^{1}_{k},X^{2}_{k},\ldots
X^{h}_{k}\}\in Y$ is referred to as {\it local maximal k-mer} if
for any sub-maximal maxmer $y_{k+p}=\{X^{1}_{k+p},\ldots
X^{b}_{k+p}\}$, where $p\ge 1$ $\exists X^{i}_{k}\in y_{k}$ such that
$X^{i}_{k}\bar{\subset} X^{j}_{k+p}\in y_{k+p}$, $j=1\ldots b$.

{\bf Definition 4}. A local maximal k-mer is referred to as a {\it super maximal
k-mer} if the conditions of definition 3 are valid for all
$X^{i}_{k}\in y_{k}$.

In the example above the subsequence $ACAGGTA$ represents a supermaximal
$7$-mer, while three sequences $GGTTA$ yield a local maxmer, as only the first
such sequence can not be extended while two other sequences can be extended to
supermaximal maxmer $ACAGGTA$.

It is seen that relations of maximal exact matches and supermaximal and
local maximal maxmers are not straightforward. One may roughly say that
the set of all supermaximal repeats would be a subset of all maximal exact matches. However
insignificant deviations from this inclusion can appear because we define
maximal exact matches as {\it pairs} of elements while supermaxmers even for DNA
sequences can consist of three sequences; but such supermaxmers are so rare that
their influence is negligible and in a zeroth approximation we can rely on the
relation indicated above. The connections to local maxmers are more subtle: from
the example above it is clear that maximal exact matches are often ``chosen'' as
pairs from local maxmers containing many sequences. Though it is correct that
supermaximal and local maxmers suggest more non-trivial division of repeats in
the chromosome, maximal exact matches as we defined them above provide an
independent measure of non-local correlations in DNA.

\section{Appendix 2. To the definition of length distribution.}

\noindent
{\bf\S 5.}
Based on the previous definitions of various repeats we provide more rigorous
treatment of the length distribution.

{\bf Definition 5}. The number of $X^{j}_{k}$ containing in sub-maximal k-mer
is referred to as {\it index} of the sub-maximal k-mer with respect to the set
$X$ and is denoted by $\In_{X}(y_{k})$. 

Thus $\In_{X}(y_{k})=h$ (cf. definition 1). This obviously would
correspond to introducing some indicator function on the set $Y$\footnote{There
may exist sensible definitions of index different from definition 5, from which
we mention the following: if $y_{k}$ is a submaximal k-mer from def. 1 with
$h>1$, then $\In_{X}(y_{k})=1$ for any $h$. One may say that in definition 5 the index counts 'occurrences' of a sequence $y_{k}$ in $X$,
while in the last definition the number of sub-maximal $k$-mers is counted;
this terminology is developed in \cite{taillefer_miller}}. According to the 
definition 1, $\min_{y\in Y}\In_{X}(y_{k})=2$. In addition, the function
$\In_{X}(y_{k})$ is non-negative and finite-valued. If the element $y_{k}$ is not a sub-maximal k-mer, then we put $\In_{X}(y_{k})=0$.
The index is defined similarly for all types of repeats introduced in \S\S 3,4.

\noindent
{\bf\S 6.}
Let us introduce an
equivalence relation on $Y$. Two elements of $Y$ are equivalent 
if they are both sub-maximal $k$-mers wrt. $X$. Thus, the
set $Y$ is partitioned into classes of equivalent elements.
The set obtained by means of factorization of $Y$ with respect to this
equivalence relation is denoted by $Y^{X}_{F}$. Thus, each element
$y^{F}_{k}\in Y^{X}_{F}$  consists of all sequences $y\in Y$ of $k$ elements
intersecting to $X$ and included to some (sub)maximal $k$-mer.

The notion of index is easily redefined for arbitrary equivalence classes (not
only for sub-maximal k-mer but for maximal exact matches or supermaxmers). These definitions are straightforward and we omit them.

{\bf Definition 5'}. If $y^{(1)}_{k}, y^{(2)}_{k}\ldots, y^{(p)}_{k}\in Y$ are
equivalent with respect to the equivalence relation $F$, then the index of the
corresponding element $y^{F}_{k}\in Y^{X}_{F}$ is given by
\begin{equation}
\label{index}
\In(y^{F}_{k})=\sum_{i=1}^{p}\In_{X}(y^{(i)}_{k}).
\end{equation}

\vspace{0.5cm}
\noindent
{\bf\S 7.}
{\bf Example.} We can consider the notion of index in application specifically
to supermaxmers. In this case the configuration space is $S=\{A,
T, C, G\}$ and  supermaximal maxmers can contain $2, 3$ or $4$
sequences\footnote{in binary case, only two sequences. The number of
supermaxmers with $3$ or $4$ sequences is negligible compared to those with
two sequences.}.
Thus, according to definition 5 the corresponding indexes are equal to $2, 3$
and $4$. The space $Y^{X}_{F}$ is obtained by establishing the equivalence of all supermaxmers,
which have the same number of elements.

 The complete number of elements
containing in $y^{F}_{k}\in Y^{X}_{F}$ is given by (\ref{index}). As each $y\in Y$ belongs to at least one $y^{F}_{k}$, then $Y$ is partitioned
into equivalence classes with respect to supermaximal sequences. Consequently
$\In(y^{F}_{k})$ can be computed for any $y^{F}_{k}$. Then we can
introduce the following definition.

{\bf Definition 6}. The function $n(k) = \In(y^{F}_{k})$, $y^{F}_{k}\in
Y^{X}_{F}$, $k=1,2,\ldots$ is referred to as {\it empirical length distribution}
on $Y$ wrt. $X$.

\noindent
{\bf\S 8.}
It is important to
notice that the equivalence relation is constructed for studying some
correlation properties of m-ary sequences, e.g., genomes, which do not depend on
a concrete structure or content of these sequences but which would
incorporate physical length as one of the governing parameters. In this context it should
be understood that there are many other ways to construct an equivalence
relation or, in physical terms, {\it coarse graining} on $Y$. However, these
definitions typically neglect the physical length. The simplest way is to
include only supermaximal $k$-mers and neglect local ones. To give a less obvious and exotic example we may say that two elements of $Y$ are equivalent
if, provided that configuration space is $S=\{0,1\}$, they
contain equal fractions of 1s.  This is especially easy to envisage for binary sequences but 
also may be reasonable for arbitrary m-ary sequences.
In part, the similar construction was applied in \cite{lee} to produce so
called $k$ spectra of genomes. As genetic 'alphabet'
consists of $4$ letters the authors consider $k$-mers with respect to the
fraction of (A+T) content. In our terms that means introducing a different
equivalence relation on the set $Y$ than one mentioned above. On the other
hand, we may consider the trivial equivalence relation when any $y\in Y$ is
equivalent only to itself.
This situation is ubiquitously exploited, e.g., in genomics where one can take
a specific ``functional" sequence and ask whether its copies are found in
different genomes. In this situation
the content of the sequence is not eliminated because the assumed
functionality implies that any nucleotide may be important. 
The interesting example of manipulations with this limiting case of self-equivalency is given in \cite{lee}.

\pagebreak
\widetext
\begin{center}
\textbf{\large Supplemental Materials: On a chain of fragmentation equations for
duplication-mutation dynamics in DNA sequences}
\end{center}
\setcounter{equation}{0}
\setcounter{figure}{0}
\setcounter{table}{0}
\setcounter{page}{1}
\makeatletter
\renewcommand{\theequation}{S\arabic{equation}}
\renewcommand{\thefigure}{S\arabic{figure}}
\renewcommand{\bibnumfmt}[1]{[S#1]}
\renewcommand{\citenumfont}[1]{S#1}

\section{Supplemental figures}

\begin{figure}[h]
\hspace{-1.5cm}
\includegraphics[width=400pt, height=300pt]{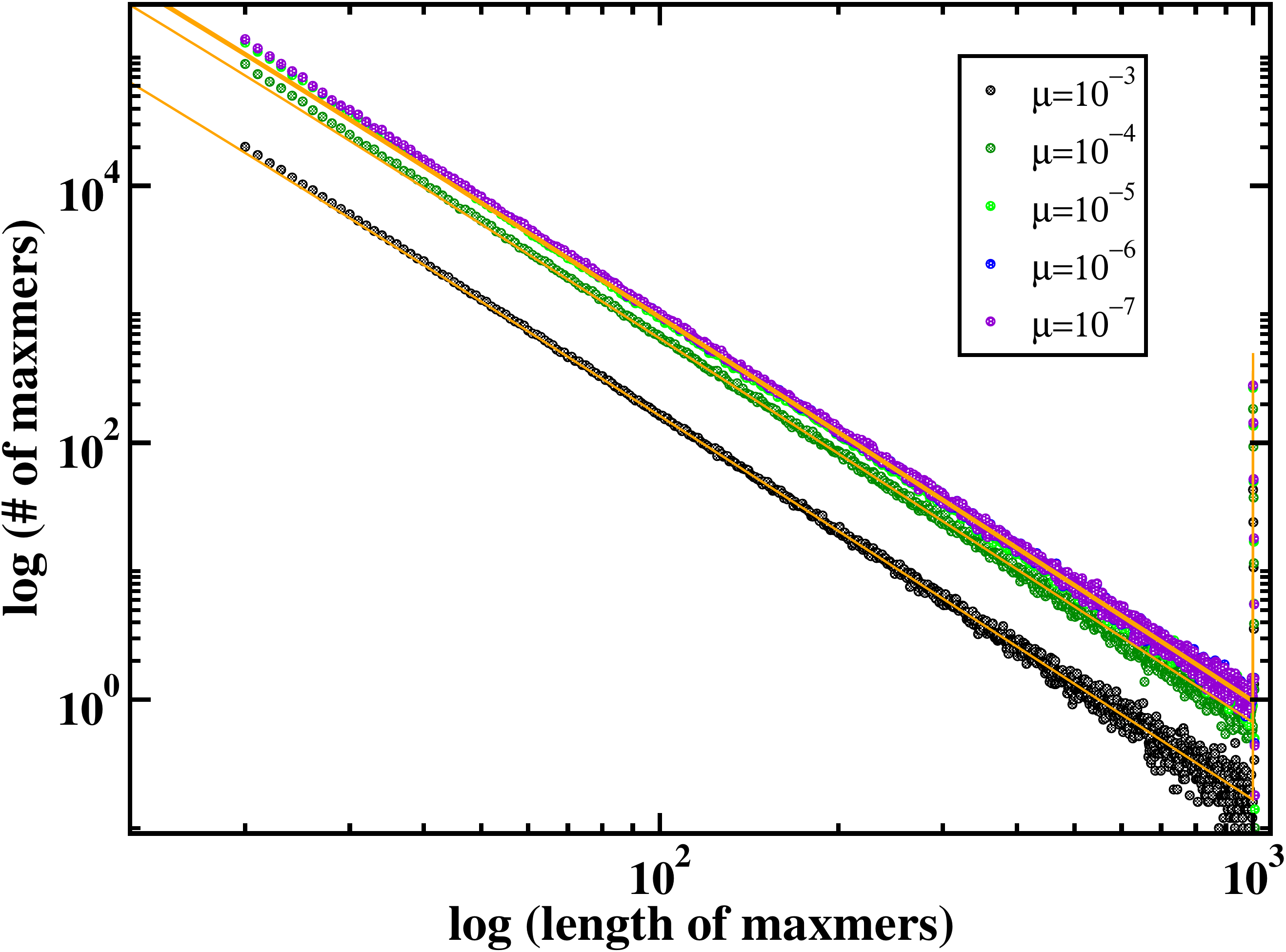}
\caption{Comparisons of simulations with solutions of equation (4) of the
main text. The parameters are: $L=10^{6}$, $D=10^{3}$, $\lambda=10^{-1}$.
Empirical length distributions were computed with the same switches of {\it
mummer} as indicated in the caption for figure 1 of the main text. The
distributions were averaged over $100$ realizations.}
\end{figure}

\begin{figure}
\hspace{-1.5cm}
\includegraphics[width=400pt, height=300pt]{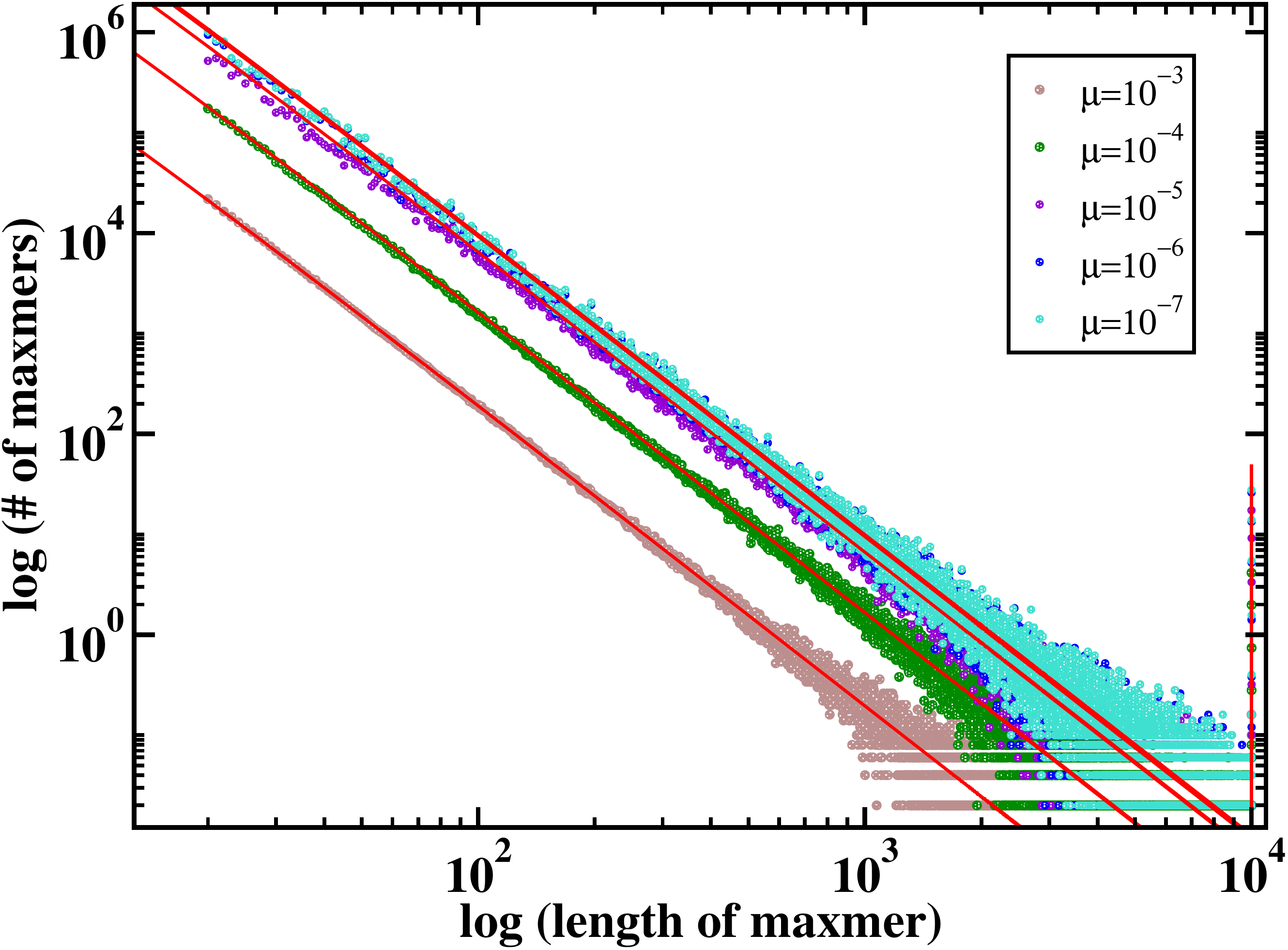}    
\caption{Parameters of the model are: $L=10^{6}$, $D=10^{4}$,
$\lambda=10^{-1}$. All other parameters and options are the same as in 
figure 1 of the main text and supplemental figure 1.}
\end{figure} 


\begin{figure}
\hspace{-1.5cm}
\includegraphics[width=400pt, height=300pt]{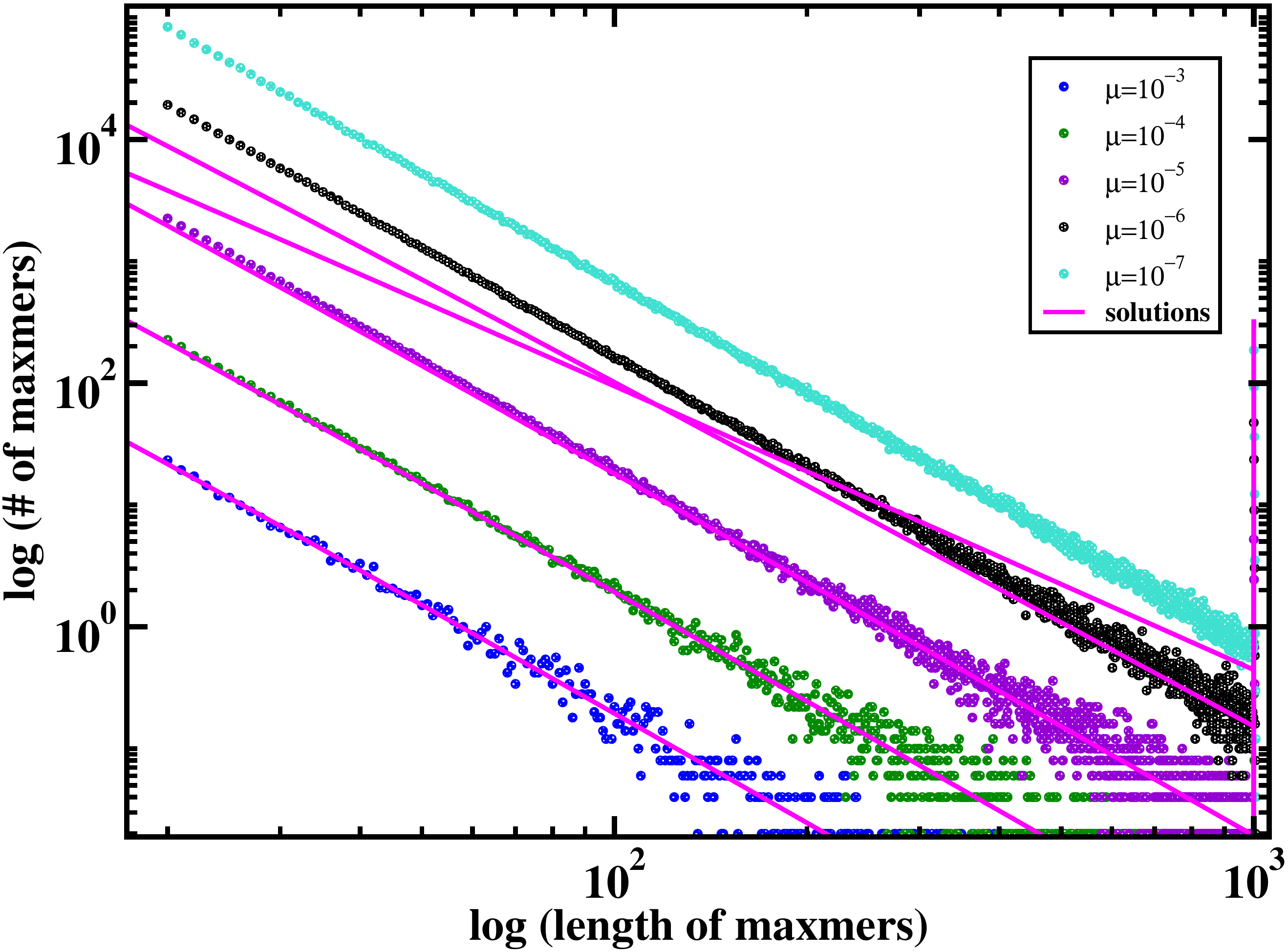}
\caption{Length distributions obtained with duplication-mutation dynamics using {\it mummer} with
the parameters {\it -n -b -l 20}. Parameters of the model are: $L=10^{6}$, $D=10^{3}$,
$\lambda=10^{-4}$ and correspond to those indicated in the fig. 1 of the main text. Magenta curves represent
the solutions of the equation (2) of the main text.}
\end{figure}



\begin{thebibliography} {99}
\bibitem{footnote3-1}{both types of repeats include a condition of maximality
of exact matches; this condition, however, turns out to be insufficient to
obtain similar length distributions.}
\bibitem{footnote5}{Resolution of this issue will become possible when the
software for computation of local maxmers will have been publicly available. The
task of local maxmers computation and analysis is nontrivial and is out of the scope of
the paper.}
\end{thebibliography}
\end{document}